\documentclass[multphys,vecphys]{svmult}

\usepackage{makeidx}         % allows index generation
\usepackage{graphicx}        % standard LaTeX graphics tool
\usepackage{multicol}        % used for the two-column index
\usepackage[bottom]{footmisc}% places footnotes at page bottom
\makeindex             % used for the subject index
\begin{document}

\title*{A Pan-Spectral Method of Abundance Determination}
\author{A. Sapar\and A. Aret\and L. Sapar\and R. Poolam{\"a}e}
\institute{Tartu Observatory, 61602 T{\~o}ravere, Estonia
\texttt{sapar@aai.ee}}

\maketitle

\begin{abstract}
We propose a new method for determination of element abundances in stellar atmospheres aimed for
the automatic processing of high-quality stellar spectra.
The pan-spectral method is based on weighted cumulative line-widths
\mbox{$Q_\lambda=\int_{\lambda_0}^\lambda{\left|\frac{dR_\lambda}{dZ}\right|}(1-R_\lambda)d\lambda$,}
where $R_\lambda$ is residual flux and $Z$ is abundance of studied element.
Difference in quantities $Q_\lambda$ found from synthetic and observed spectra gives a correction to the initial abundance.
Final abundances are then found by rapidly converging iterations. Calculations can be made for
many elements simultaneously and do not demand supercomputers.
\end{abstract}

\section{Description of the method}
\label{sap:sec1}
Essential developments in observational high-precision and high-resolution spectroscopy, and
fast-growing computing facilities stimulate to seek new automatic methods of analysis of stellar spectra.
We recently started to develop a pan-spectral (or broad-band) method for determination of
abundances of chemical elements and their isotopes in stellar atmospheres and
to compose a software necessary for its application.

The method is based on weighted cumulative line-widths $Q_\lambda$ defined as
\begin{equation}
      Q_\lambda=\int_{\lambda_0}^\lambda{\left|dR_\lambda \over dZ\right|}(1-R_\lambda)d\lambda~,
\label{sap:Q}
\end{equation}
where $R_\lambda$ is the residual flux (intensity) and $Z=\log(N_{elem}/N_{tot})$ is the abundance of studied element or isotope.

Integral (\ref{sap:Q}) can be easily reduced to the equivalent
width of a spectral line by omitting the derivative $\left|dR_\lambda
/ dZ\right|$ and integrating only over wavelengths of this line.
We call the quantity $Q_\lambda$ cumulative because integral is
taken over all lines starting from some initial wavelength to the
current wavelength. And we call it weighted because the derivative
of residual flux $R_{\lambda}$ with respect to abundance $Z$ can
be treated as a weight function. This derivative automatically
excludes spectral regions insensitive to changes of the abundance of studied element and
gives a large contribution in the most sensitive regions, i.~e. in the centers of non-saturated
lines and in the steep wings of strong lines of this element.

Weighted cumulative widths $Q_\lambda$ for both observed and
synthetic spectra are calculated using the derivative of
$R_{\lambda}$ found from the synthetic spectra. For small
abundance changes the derivative can be replaced by differences,
thus
\begin{equation}
      \frac{dR_\lambda}{dZ}= \frac{R^+_\lambda - R^-_\lambda}{Z^+ - Z^-}~,
\label{sap:deriv}
\end{equation}
where flux $R^+_\lambda$ corresponds to abundance  $Z^+ = Z + \delta Z$ and
flux $R^-_\lambda$ corresponds to abundance  $Z^- = Z - \delta Z$.
Correction to the abundance of studied element $Z$ can then be found by formula
\begin{equation}
      \Delta Z={(Z^+-Z^-)(Q_\lambda^{obs}-Q_\lambda^{syn}) \over  Q_\lambda^+-Q_\lambda^-}~,
\label{sap:dZ}
\end{equation}
where $Q_\lambda^{obs}$ is found from the observed spectrum and $Q_\lambda^{syn}$ from the
synthetic spectrum with initial (guessed) abundance $Z$. Quantities $Q_\lambda^+$ and $Q_\lambda^-$ correspond
to synthetic spectra with abundances $Z^+$ and $Z^-$, respectively.

Abundance of element can be found iteratively using the corrections $\Delta Z$,
the number of iterations needed depends on difference between initial and best-fit abundance values.
Abundance determined by such a procedure is a best fit for all lines of the element,
diminishing thus an influence of possible atomic data errors. We would like to point out that here
the contribution of blended lines is duly taken into account.
Abundances can be found simultaneously for many elements.
Our calculations show that the method works well for line-rich elements, but for line-poor elements
further refinements are necessary.

The method can be used also for determination of the effective temperature and gravity, using as
the weight function the derivative of residual flux with respect to $T_{\rm eff}$ or $\log g$, respectively.

\section{Testing the method}
\label{sap:sec2}

\subsection{Observational Data}
\label{sap:sec2.1}

We applied the weighted cumulative width method to determine
element abundances in the atmosphere of the chemically peculiar star HD 175640.
High resolution spectral atlas of this star, covering the 3~040 -- 10~000 \AA ~region,
was recently published by Castelli and Hubrig \cite{sap:castelli}.
The spectrum with resolution 90~000 -- 110~000
and signal-to-noise ratio 200 -- 400 was obtained at ESO 8 m UT2 telescope
with the UV-Visual Echelle Spectrograph (UVES). Castelli and Hubrig \cite{sap:castelli} also computed
a synthetic spectrum for the whole observed interval with the SYNTHE code \cite{sapar_kursyn} and determined
abundances for 48 ions of 39 elements  by the visual best fit of synthetic and observational spectra of selected lines.
The model atmosphere accepted was an ATLAS12 model \cite{sapar_kur12}
with parameters T$_{\rm eff}$ = 12~000 K, $\log g$ = 3.95, $\xi$=~0~km~s$^{-1}$.
Improved and extended Kurucz atomic line list used in computations is available at F.~Castelli website
(\texttt{http://wwwuser.oat.ts.astro.it/castelli/stars.html}).

We had to ignore most of the IR spectrum due to telluric lines and also some spectral regions
with weak overlap or gaps between adjoined echelle spectral orders.

\subsection{Synthetic spectrum}
\label{sap:sec2.2} In computations  of  synthetic spectra we
based on the Kurucz ATLAS9~\cite{sapar_kuratlas} model
atmospheres and on the line lists and initial element abundances
provided by Castelli and Hubrig~\cite{sap:castelli}.
Computations  of the synthetic spectra were carried out
 with program SMART composed by our team
during last decade~\cite{sap:tueb}. SMART is a compact and simple software and does not
demand large computer facilities.
We used the same spectral resolution (500 000) as Castelli and Hubrig in
computations with SYNTHE \cite{sapar_kursyn}.
Gauss broadening and the noise cutoff parameters were found to achieve best fit of
observed and synthetic spectra.
Obtained synthetic spectrum agrees reasonably well with the observed one.

\subsection{Application of the pan-spectral method}
\label{sap:sec2.3} To estimate the domain of applicability of the
method, we obtained abundances of 12 line-rich elements
(Mn, Ti, Cr, Fe, Y, Mg, Si, Ca, O, S, Yb, Ni) in the atmosphere of HD 175640.
Regions of Balmer and Paschen continua were studied separately because
abundances of studied metals turned out to be larger in outer layers
of atmosphere (UV region) than in inner layers (visual region).

\begin{figure}[h]
\centering \leavevmode
  \includegraphics[bb=58 50 400 302, clip, width=0.49\textwidth]{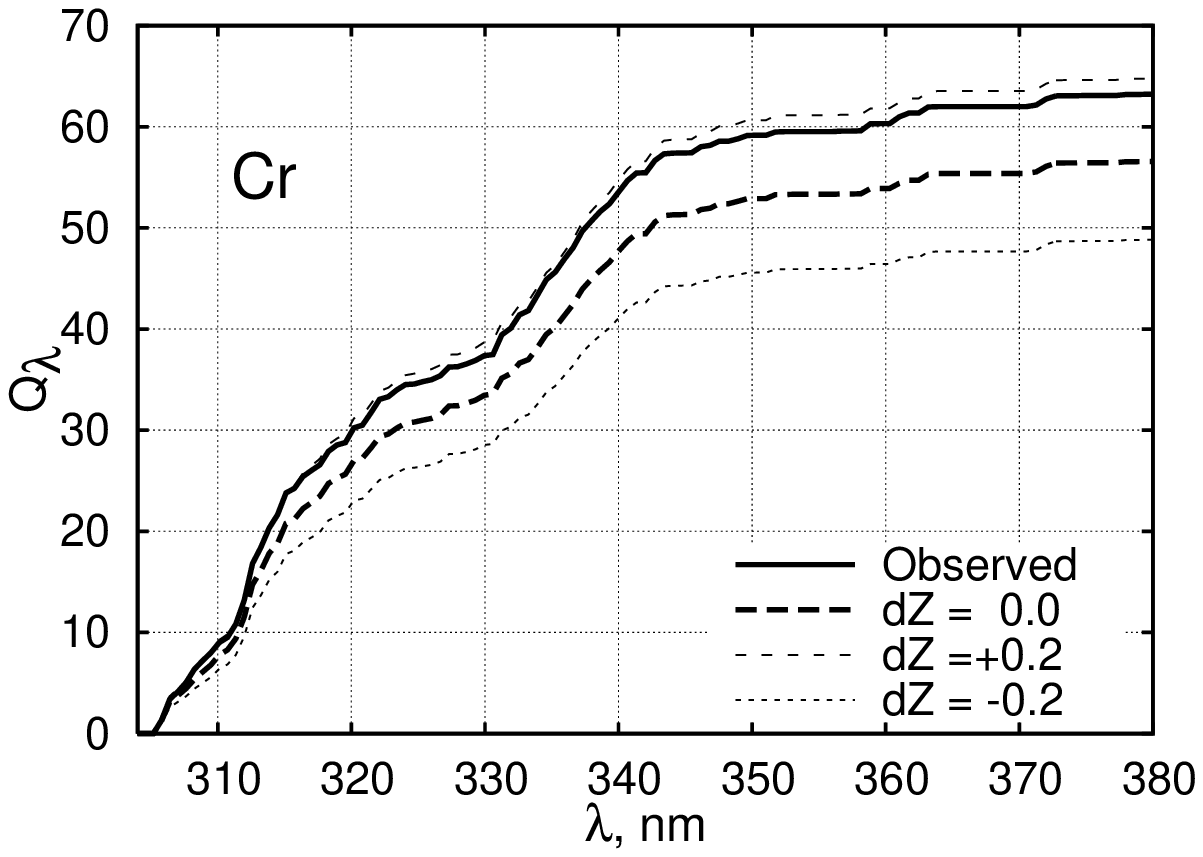} \hfil
  \includegraphics[bb=58 50 400 302, clip, width=0.49\textwidth]{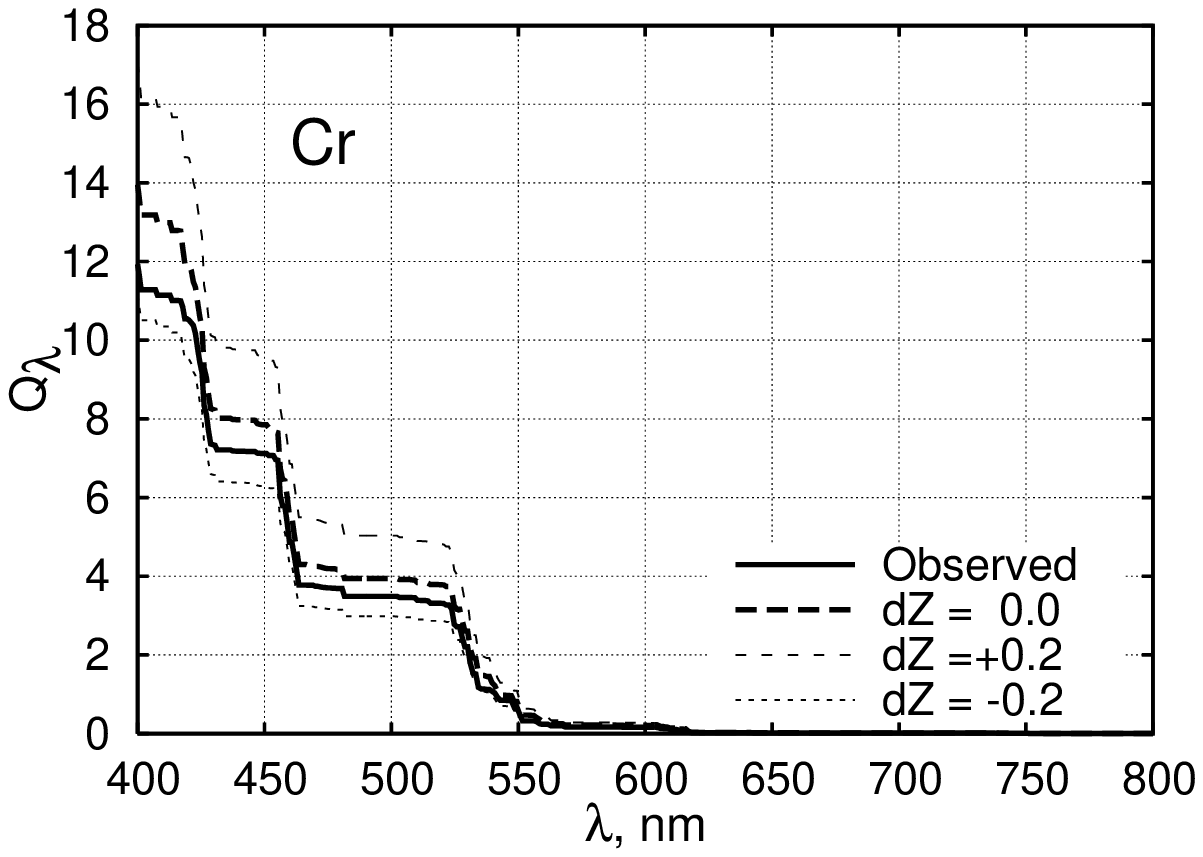}
  \caption{Weighted cumulative width $Q_\lambda$ for Cr.}
   \label{sap:Qlam}
\end{figure}

We started with abundances derived by Castelli and
Hubrig \cite{sap:castelli} and iteratively found corrections to
these values.
Abundances of Si, O and S from UV spectrum have not been found
since there were only few weak lines of these elements in UV region.
Final abundances were found with only two iterations since necessary corrections were small.
Weighted cumulative widths $Q_\lambda$ for chromium computed with initial abundance
are presented in Fig. \ref{sap:Qlam}.
Corresponding corrections to Cr abundance are +0.16 dex in UV and -0.14 dex in visual region.

Obtained abundances of the studied elements and comparison with results of
Castelli and Hubrig~\cite{sap:castelli} are presented in Table \ref{sap:Zcorrections}.
Our results differ somewhat from the abundances obtained by Castelli and Hubrig.
However, the differences in obtained element abundances may partly be
caused by use of different spectral synthesis programs, not only by difference of used methods.
This is particularly true for S and Si. Also the difference of used model atmospheres should be mentioned.

We conclude that the proposed method is promising
 for analysis of chemical composition and other main parameters
of stellar atmospheres.

\begin{table}[h]
\centering
\caption{Element abundances $[N/N_{\odot}]$ for HD175640}
\label{sap:Zcorrections}
\begin{tabular}{l|rrr|rrr|rr}
\hline
 & \multicolumn{6}{c|}{Castelli \& Hubrig (2004)} & \multicolumn{2}{c}{Present study}  \\
 \hline
 & \multicolumn{3}{c|}{UV}& \multicolumn{3}{c|}{Vis} & UV & Vis \\
 & ion I & ion II & ion III & ion I & ion II & ion III &  & \\
\hline
Mn & +2.47 & +2.40 & -- & +2.46 & -- & -- & +2.75 & +2.57 \\
Ti & -- & +1.43 & -- & -- & +1.30 & -- & +1.62 & +1.31 \\
Cr & +1.19 & +1.03 & -- & +1.13 & +0.96 & -- & +1.24 & +0.96 \\
Fe & -0.36 & -- & -- & -0.21 & -0.30 & -- & -0.23 & -0.33 \\
 Y & -- & +3.38 & -- & -- & +3.01 & -- & +3.26 & +3.04 \\
Mg & -- &  &  & +0.18 & -0.25 & -- & -0.19 & -0.09 \\
Si & -- &  &  & -- & -0.23 & -0.09 & -- & +0.21 \\
Ca & -- & -0.15 & -- & +0.42 & +0.06 & -- & +0.10 & +0.24 \\
 O & -- &  &  & +0.03 & -- & -- & -- & +0.10 \\
 S & -- &  &  & -- & -0.41 & -- & -- & +0.03 \\
Yb & -- & +3.14 & +3.66 & -- & +2.76 & -- & +3.64 & +2.79 \\
Ni & -- & -0.22 & -- & -- & -0.35 & -- & -0.12 & -0.43 \\
\hline
\end{tabular}
\end{table}

\medskip
We are grateful to Estonian Science Foundation for grant  ETF 6105.


\begin{thebibliography}{99.}

\bibitem{sap:castelli}  F. Castelli, S. Hubrig: A\& A \textbf{425}, 263 (2004)
\bibitem{sapar_kur12} R. Kurucz: Model Atmospheres for Individual Stars with Arbitrary Abundances.
In: \textit{The Third Conference on Faint Blue Stars},
ed by A. G. D. Philip, J. Liebert, R. A. Saffer (L. Davis Press, Schenectady 1997) p 33
\bibitem{sapar_kuratlas} R. Kurucz:
%ATLAS9 Stellar Atmosphere Programs and 2 km s$^{-1}$ grid,
CD-ROM, No. 13 (1993)
\bibitem{sapar_kursyn} R. Kurucz:
%SYNTHE Spectrum Synthesis Programs and Line Data,
CD-ROM, No. 18 (1993)
\bibitem{sap:tueb} A. Sapar, R. Poolam{\"a}e: ``SMART'': A Compact and Handy FORTRAN Code for the Physics of Stellar Atmospheres.
In: \textit{ASP Conf. Ser. 288: Stellar Atmosphere Modeling}, ed by  I. Hubeny, D. Mihalas, K. Werner
(Astronomical Society of the Pacific, San Francisco 2003) p 95
\end{thebibliography}
\end{document}